\documentclass[twocolumn,superscriptaddress,nolongbibliography,amsmath,amssymb,aps,prb]{revtex4-2}
\usepackage[colorlinks=true,urlcolor=blue,citecolor=blue,linkcolor=blue]{hyperref}
\usepackage{txfonts}
\usepackage{appendix}
\usepackage{graphicx}
\usepackage{subcaption}
\usepackage{tikz}
\usepackage{dcolumn}
\usepackage{bm}
\usepackage{color}
\begin{document}
\title{Bounding free energy difference with flow matching}
\author{Lu Zhao}
\affiliation{Institute of Physics, Chinese Academy of Sciences, Beijing 100190, China}
\affiliation{University of Chinese Academy of Sciences, Beijing 100049, China}
\author{Lei Wang}
\email{wanglei@iphy.ac.cn}
\affiliation{Institute of Physics, Chinese Academy of Sciences, Beijing 100190, China}
\affiliation{Songshan Lake Materials Laboratory, Dongguan, Guangdong 523808, China}
\date{\today}
\begin{abstract}
This paper introduces a method for computing the Helmholtz free energy using the flow matching technique. Unlike previous work that utilized flow-based models for variational free energy calculations, this method provides bounds for free energy estimation based on targeted free energy perturbation, by performing calculations on samples from both ends of the mapping. We demonstrate applications of the present method by estimating the free energy of the classical Coulomb gas in a harmonic trap.
\end{abstract}
\maketitle

\section{\label{sec:intro}Introduction}

Free energy calculations hold significant importance in statistical physics, chemistry and biology. By evaluating the free energy landscape, one can gain a quantitative understanding of thermodynamic properties. These calculations enable prediction, optimization, and a deeper comprehension of complex phenomena in diverse scientific fields~\cite{kollman1993free}.

Although molecular dynamics and Monte Carlo simulations allow for large-scale and high-precision sampling, the computation of the partition function poses a formidable challenge, rendering the calculation of free energy using these methods an arduous endeavor~\cite{frenkel2023understanding,tuckerman2023statistical}. In general, to compute the free energy, one typically embarks upon the quest of identifying a reference system with a known free energy, thereby establishing a connection between the target system and the reference system. Based on this roadmap, scientists have developed two classes of computational methods. In one class of methods, there is an assumption that the system under investigation remains in thermodynamic equilibrium or, at least, experiences slow temporal changes. The difference in free energy between two systems corresponds to the work done along the switching pathway. Researchers calculate this difference by performing thermodynamic integration~\cite{frenkel2023understanding} through a series of simulations along the pathway. The second category of methods is based on the non-equilibrium equality proposed by Jarzynski~\cite{jarzynski1997nonequilibrium}. The free energy evaluation using Annealed Importance Sampling~\cite{neal2001annealed} is essentially equivalent to the Jarzynski equality. On the other hand, Free Energy Perturbation~\cite{zwanzig1954high} (FEP) can be seen as an extreme form in the non-equilibrium regime when the time approaches zero.

The development of FEP has led to the targeted free energy perturbation~\cite{jarzynski2002targeted,hahn2009using} (TFEP) method. Given an invertible mapping, TFEP provides a bound on free energy as well as a unique free energy estimator by separately computing the generalized work on both sides. However, the true power of the TFEP method is challenging to unleash using manually designed mapping expressions, as the selection of the mapping is crucial. Nevertheless, with the explosive development in the field of deep learning, a method for learning invertible mappings called normalizing flow~\cite{tabak2010density,dinh2014nice,rezende2015variational,papamakarios2021normalizing} has emerged. Leveraging the advancements in flow-based models, we can enable the computer to autonomously discover a mapping with sufficient overlap. While it is beneficial to let the computer train and select the mapping, incorporating certain human considerations can also be advantageous.

In the previous work~\cite{wirnsberger2020targeted, wirnsberger2022normalizing, caselle2022stochastic} that combined discrete flow-based models with TFEP, however, designing the structure of the flow to incorporate symmetry into the reference distribution was a challenging task. In physical systems, symmetry plays a crucial role, and imbuing the reference distribution with such considerations to narrow down its variational space can be valuable. Though continuous normalizing flows~\cite{zhang2018monge,chen2018neural} (CNFs), which employ the construction of a velocity field to establish an invertible mapping between two spaces through ordinary differential equation (ODE) integration, enable us to conveniently incorporate symmetry into the constructed velocity field, their training method based on ODE integration results in suboptimal computational efficiency and accuracy. The advancement in the field of normalizing flows has introduced the flow matching method~\cite{lipman2022flow,liu2022flow,albergo2022building}. Flow matching not only retains the concise and elegant characteristics of CNFs, but also demonstrates remarkable precision and efficiency due to circumventing the need for ODE integration during training, and has been showcased in the fields related to free energy~\cite{klein2023equivariant}.

It is worth noting that in many previous works using flow-based models for physical problems, researchers trained the models variationally without heavy reliance on data, thus enabling the widespread adoption of this elegant method across various domains in recent years, such as lattice models~\cite{li2018neural,nicoli2020asymptotically}, molecules~\cite{li2020neural} and hydrogen atoms~\cite{xie2022ab,xie2023m,xie2023deep}. However, variational free energy methods, in principle, only provide an upper bound estimation of free energy and do not achieve exact computation.

In our study, we demonstrated the application of the targeted free energy perturbation with flow matching to the classical Coulomb gas in a harmonic trap~\cite{bolton1993classical}, providing bounds for the free energy. Notably, By plotting a frequency histogram of "forward work" and "reverse work", we can observe that the overlap of the two frequencies occurs precisely within the bounds.

\section{\label{sec:meth}Methods}

\subsection{\label{sec:fe}Free energy estimation}

We employ $A$ and $B$ to denote two thermodynamic equilibrium states, with their respective densities being $\rho_A(x) = \mathrm{e}^{-\beta H_A(x)}/ Z_A$ and $\rho_B(x) = \mathrm{e}^{-\beta H_B(x)}/ Z_B$, where $x$ denotes a point in the configuration space, $Z=\int \mathrm{d}x\mathrm{exp}[-\beta H(x)]$ is the partition function and $\beta$ is the inverse temperature.

Given an invertible mapping $\mathcal{M}: A \to A^\prime$, we can map $A$ to a new state $A^\prime$, transforming configurations $x$ sampled from $A$ into new configurations $y = \mathcal{M}(x)$. Similarly, we also consider the reverse case where configurations are drawn from $B$ and mapped to $B^\prime$ via the inverse $\mathcal{M}^{-1}: B \to B^\prime$. We refer to this pair of prescriptions as the "forward" and "reverse" processes, while employing the symbols $\to$ and $\gets$ to represent them, respectively. For each process, we denote generalized energy differences as
\begin{equation}
    \label{eq:Phi}
    \left\{
        \begin{aligned}
        \Phi_\to(x) &= H_B(\mathcal{M}(x)) - H_A(x) -\beta^{-1}\mathrm{log}|J_\mathcal{M}(x)| \\
        \Phi_\gets(x) &= H_A(\mathcal{M}^{-1}(x)) - H_B(x) -\beta^{-1}\mathrm{log}|J_{\mathcal{M}^{-1}}(x)|,
        \end{aligned}
    \right.
\end{equation}
where $J_\mathcal{M}$ and $J_{\mathcal{M}^{-1}}$ are the Jacobian determinants associated with the mappings. We refer to the generalized energy differences as "forward work" and "reverse work" respectively.

By applying the Kullback-Leibler divergence to $A^\prime$ and $B$ and considering the non-negative property, we can obtain 
\begin{equation}
    \int^\infty_{-\infty} \rho_{A^\prime}(x)\mathrm{ln} \left( \frac{\rho_{A^\prime}(x)}{\rho_B(x)}\right) dx \geq 0,
\end{equation}
where the inequalities can be derived:
\begin{equation}
    \label{eq:vfe}
    \Delta F \leq \left\langle \Phi_\to \right\rangle _A.
\end{equation}
Here, $\Delta F = -\frac{1}{\beta}(\mathrm{ln}Z_B - \mathrm{ln}Z_A)$ is defined as the free energy difference between $A$ and $B$. The symbol $\left\langle \cdot \right\rangle$ represents the ensemble average under the specified state. Equation~\ref{eq:vfe} represents the principle of variational free energy, providing an upper bound on the free energy difference by computing the ensemble average of the "forward work" $\Phi_\to$ under the state $A$. When the free energy of one state is known (e.g. Gaussian), this becomes a way to estimate the absolute free energy of the other state. By interchanging the probability distributions within the Kullback-Leibler divergence, we can derive the following constraints:
\begin{equation}
    \label{eq:bound}
    \left\langle - \Phi_\gets \right\rangle _B \leq \Delta F \leq \left\langle \Phi_\to \right\rangle _A.
\end{equation}
Therefore, we can establish the upper and lower bound of free energy by estimating the respective expectations of $\Phi_\to$ and $- \Phi_\gets$ on the two distributions respectively. It is evident that as the Kullback-Leibler divergence between the two distributions $\rho_{A^\prime}$ and $\rho_B$ (or, $\rho_{B^\prime}$ and $\rho_A$) approaches zero, the bounds become constricting. When the two distributions are perfectly identical, both inequalities in the equation hold true.

Moreover, The fluctuation theorem~\cite{crooks2000path,hahn2009using} exists between the two processes
\begin{equation}
    \label{eq:ft}
    \frac{p_\to(\phi)}{p_\gets(- \phi)} = \mathrm{e}^{\beta(\phi-\Delta F)}.
\end{equation}
\begin{equation}
    \label{eq:gdf}
    p_\to(\phi) = \int \delta(\phi - \Phi_\to(x))\rho_A(x)dx
\end{equation}
and
\begin{equation}
    \label{eq:gdr}
    p_\gets(\phi) = \int \delta(\phi - \Phi_\gets(x))\rho_B(x)dx
\end{equation} 
are generalized work distributions and $\delta$ is the Dirac delta function. It means when we present a probability distribution in graph, the precise value of free energy is inevitably situated at the intersection of the two probability distribution graphs.

\subsection{\label{sec:fm}Flow matching}

\begin{figure}
    \centering
    \begin{subfigure}{0.12\textwidth}
        \centering
        \includegraphics[width=\linewidth]{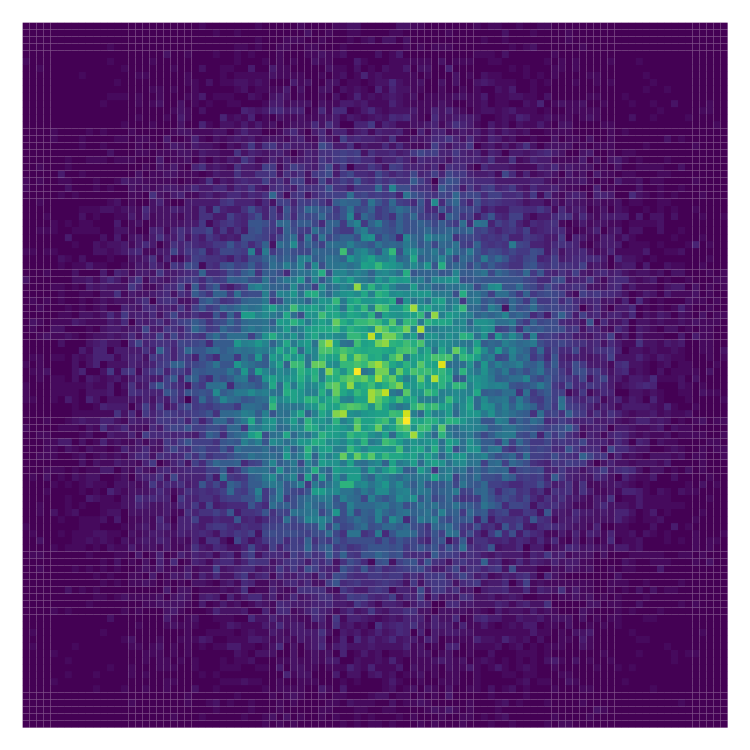}
        \caption{Density $\rho_A$}
        \label{fig:base}
    \end{subfigure}
    \hspace{4cm}
    \begin{subfigure}{0.12\textwidth}
        \centering
        \includegraphics[width=\linewidth]{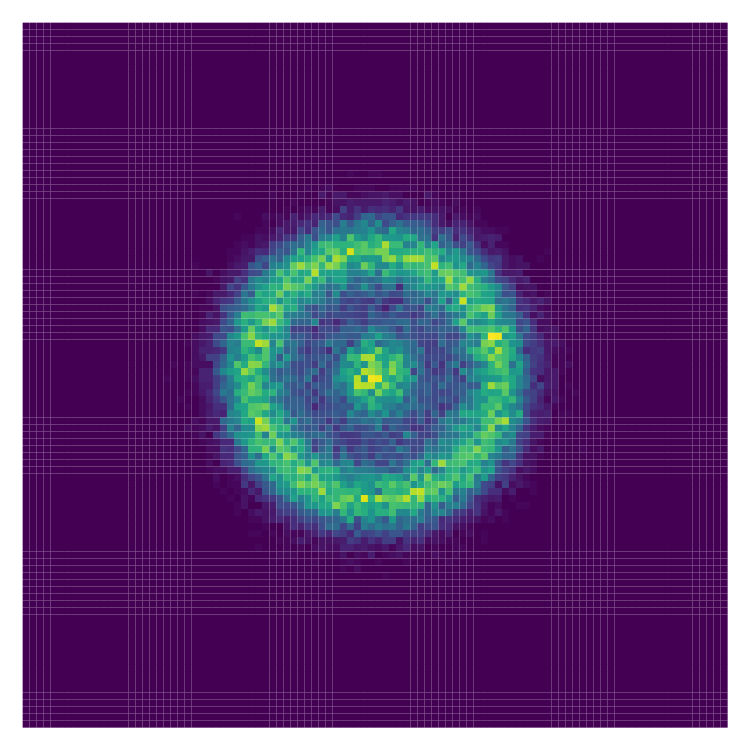}
        \caption{Density $\rho_B$}
        \label{fig:mcmc}
    \end{subfigure}
    
    \begin{tikzpicture}[overlay, remember picture]
        \draw[->] (-1.5,2.3) -- (1.5,2.3);
        \draw[->] (1.5,1.8) -- (-1.5,1.8);
        \node at (0,2.6) {$\int_{0}^{1} -\nabla\cdot v dt$};
        \node at (0,1.5) {$\int_{0}^{1}\nabla\cdot v dt$};
    \end{tikzpicture}

    \caption{The figures respectively show the probability densities of the states $A$ and $B$, which  can be connected between the two through the integration of the velocity field via Equation~\ref{eq:cnf}. In this schematic diagram, Figure~\ref{fig:base} represents a Gaussian distribution, while Figure~\ref{fig:mcmc} represents the distribution obtained through MCMC using Equation~\ref{eq:H}.}
\end{figure}
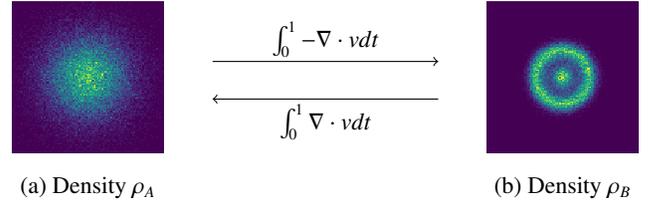

In order to make the bounds~\ref{eq:bound} narrower and the overlap between the two distributions $p_\to(\phi)$ and $p_\gets(-\phi)$ more pronounced, we require an invertible mapping $\mathcal{M}$ that allows $\rho_{A^\prime}(x)$ to accurately approximate $\rho_B(x)$. To calculate generalized energy differences $\Phi$ according to Equation~\ref{eq:Phi}, we should not only be able to sample from $\rho_{A^\prime}(x)$, but also possess knowledge of the specific values of the probability density corresponding to each sample. This, indeed, is the very essence of the normalizing flow's characteristic.

A flow model is a probability distribution $p_\mathcal{X}: \mathbb{R}_\mathcal{X}^d \to \mathbb{R}_{>0}$ defined as the pushforward of a base distribution $p_\mathcal{Z}: \mathbb{R}_\mathcal{Z}^d \to \mathbb{R}_{>0}$ through a flexible diffeomorphism $ f: \mathbb{R}_\mathcal{Z}^d \to \mathbb{R}_\mathcal{X}^d $, typically parameterized by neural networks. When points $x \in \mathbb{R}_\mathcal{X}^d$ and  $z \in \mathbb{R}_\mathcal{X}^d$ satisfy $x = f(z)$, their corresponding probability distributions satisfy $\mathrm{log}p_x(x) = \mathrm{log}p_z(z) - |\mathrm{det}J_f(z)|$.

Neural ordinary differential equations~\cite{chen2018neural} can be seen as the continuous version of residual flows. Instead of specifying a discrete sequence of hidden layers, we parameterize the derivative of the hidden state using a neural network, then the log-likelihood can be calculated as below 
\begin{equation}
    \label{eq:cnf}
    \left\{
        \begin{aligned}
            \frac{dx}{dt} &= v, \\
            \frac{d\mathrm{ln}p}{dt} &= -\nabla \cdot v.
        \end{aligned}
    \right.
\end{equation}
In order to endow $p(x)$ with a certain symmetry, specifically $p(x) = p(\mathcal{P}x)$, it can be observed that in the context of CNFs, what we aim to do is to construct a velocity field that possesses the corresponding equivariance:
\begin{equation}
    v_t(\mathcal{P}x) = \mathcal{P}v_t(x).
\end{equation}

As optimizing the parameterized velocity field, we use a brand new method called flow matching. Instead of negative log-likelihood, flow matching is minimizing the velocity directly. By introducing a time-differentiable interpolant $I_t: \mathbb{R}^d \times \mathbb{R}^d \to \mathbb{R}^d$, such that
\begin{equation*}
    I_{t=0}(x_0, x_1) = x_0 \textrm{ and } I_{t=1}(x_0, x_1) = x_1,
\end{equation*}
the velocity $v_t(x)$ that satisfies the continuity equation with the probability density $p_t(x)$ is the unique minimizer over $v_{\theta,t}(x)$ of the objective~\cite{albergo2022building}
\begin{equation}
    \label{eq:loss}
    \mathcal{G} = \mathbb{E}_{x_0} \mathbb{E}_{x_1} \| v_{\theta,t}(I_t(x_0, x_1)) -\partial_tI_t(x_0, x_1) {\|^2}.
\end{equation}
Once we have selected an interpolant that satisfies the boundary conditions, here we choose
\begin{equation}
I_{t}(x_0, x_1) = (1-t) x_0 + t x_1,
\end{equation}
and parameterized the velocity field, we can proceed with the training based on the equation~\ref{eq:loss}.

\subsection{\label{sec:arch}Architectures}

\begin{figure}
    \centering
    \includegraphics[width=0.35\textwidth]{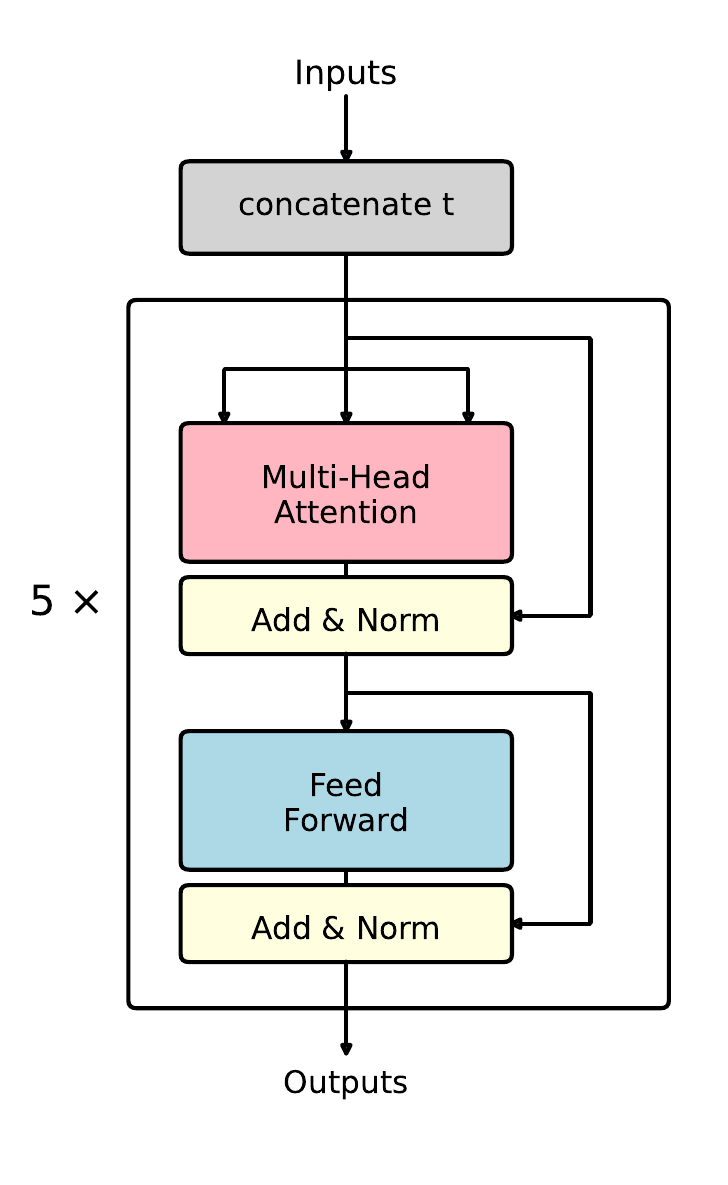}
    \caption{The network architecture for velocity field parameterization.}
    \label{fig:diag}
\end{figure}

We opt for the transformer~\cite{vaswani2017attention} architecture which satisfies the permutation symmetry of particles to parameterize a velocity field $v(x, t)$. In Figure~\ref{fig:diag}, we present the diagram of the network we used. In order to incorporate temporal dependence into the network, we input a sequence of length $n$, and we replicate the time variable $t$ $n$ times and concatenate it with each vector in the sequence, thereby augmenting its dimension to $d+1$, where $d$ denotes the dimensionality of the physical system and $n$ represents the number of particles.

Subsequently, we feed the sequence into the layer consisting of two residual connections. The first residual block comprises a multi-head attention layer, while the second one consists of a dense connection composed of two linear layers and a Gaussian error linear unit layer. After passing the sequence through the five aforementioned layers, we apply a linear transformation to the sequence, resulting in a sequence of length $n$, where each vector has a dimension of $d$. This sequence represents the velocity field that we have obtained.

\section{\label{sec:res}Results}

\begin{figure*}

    \centering
  
    \begin{subfigure}{0.95\textwidth}
        \centering
        \includegraphics[width=\textwidth]{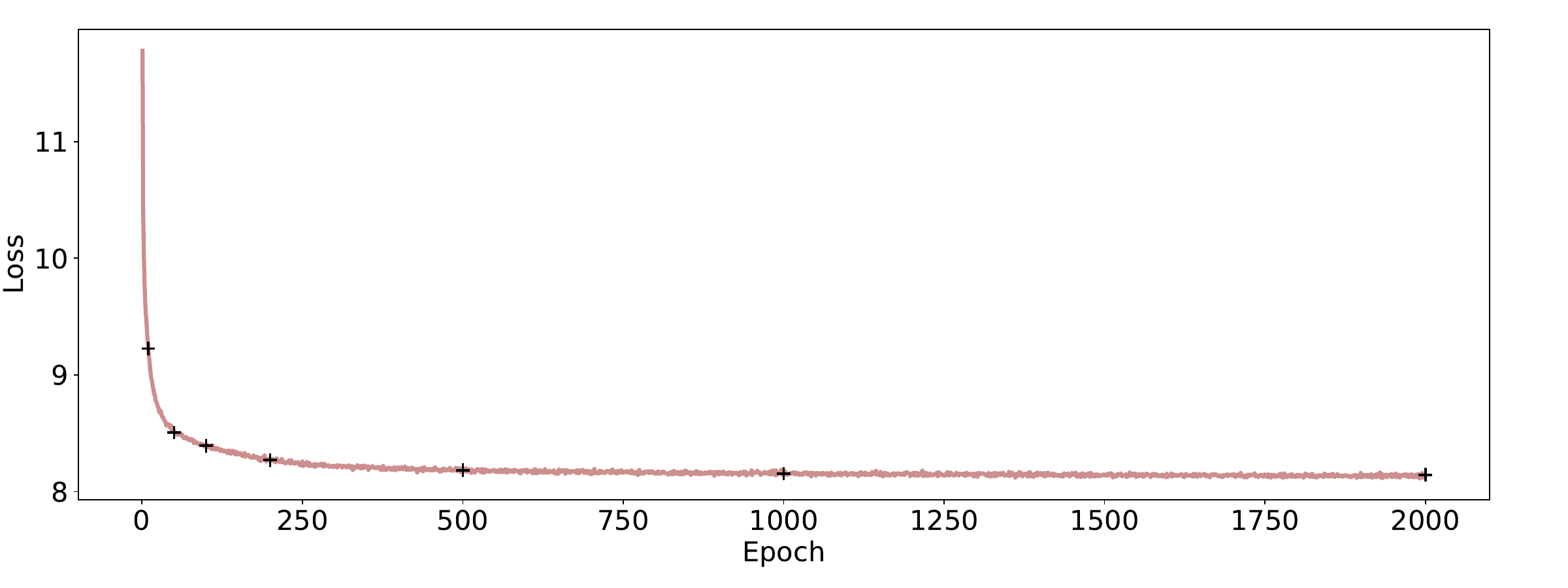}
        \begin{tikzpicture}[overlay, remember picture]
            \node at (current page.center) [xshift=0cm, yshift=9.5cm] {\includegraphics[width=0.7\textwidth]{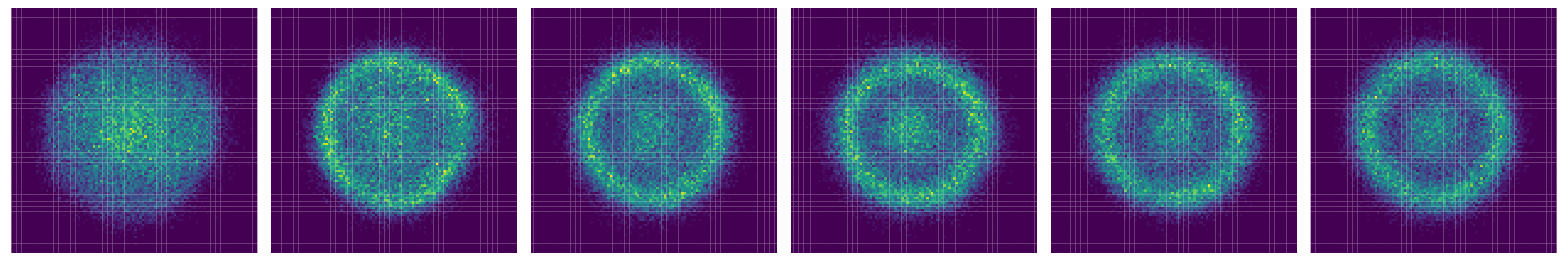}};
        \end{tikzpicture}    
        \caption{Loss}
        \label{fig:loss}
    \end{subfigure}
    \begin{subfigure}{0.95\textwidth}
        \centering
        \includegraphics[width=\textwidth]{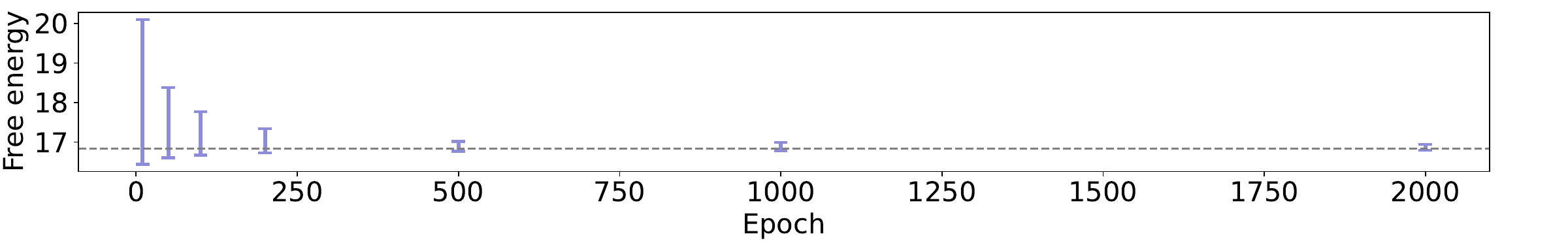}
        \caption{Free energy estimation}
        \label{fig:fe}
    \end{subfigure}

    \caption{Figure~\ref{fig:loss} depicts the evolution of the loss function~\ref{eq:loss} during training, and an array of sample distributions generated by flow-matching models at various stages of the training process. The training epochs of these models correspond to 10, 50, 100, 200, 500, 1000, and 2000 respectively, marked on the plot with black crosses. Figure~\ref{fig:fe} illustrates the upper and lower bound of free energy, derived from estimations based on $\rho_{A^\prime}(x)$ of the selected training stages, while the dashed line represents the TFEP estimation of free energy.}
    \label{fig:descent}
\end{figure*}

In this study, we have employed the flow matching method to investigate the classical Coulomb gas in a harmonic trap, where we consider the electrons as point charges and omit their kinetic energy~\cite{bolton1993classical}. The Hamiltonian reads
\begin{equation}
    \label{eq:H}
    H= \sum_{i<j} \frac{1}{|x_i - x_j|} + \sum_i  x_i^2,
\end{equation}
where $x_i$ represents the coordinate of the $i$-th electron and $\sum_i  x_i^2$ corresponds to the harmonic trapping potential eliminating the need to consider periodic boundary conditions.

We are now ready to present the compelling results obtained from our research. We conducted training on a system with a dimensionality of 2 and a particle count of 6, employing the network~\ref{sec:arch} as the trainable velocity field where the key size and the number of head are both 16 within the multi-head attention block. 

Furthermore, we consider the physical system to be in state $B$, while state $A$ is chosen to follow a Gaussian distribution whose free energy can be obtained analytically. Shown by figure~\ref{fig:loss}, as the number of epochs increases, we observe a gradual decrease in the loss function, signifying the successful optimization of our model. Figure~\ref{fig:loss} also illustrates the gradual evolution of the distribution $\rho_{A^\prime}(x)$ during the training process. It is evident that as the training deepens, the sample distribution derived from flow sampling gradually converges towards the actual distribution simulated by figure~\ref{fig:mcmc}. Simultaneously, Figure~\ref{fig:fe} shows the estimates of the upper and lower bound of free energy corresponding to state $B$ also become increasingly constrained. Despite not endowing the network with any spatial equivariance, the training results indicate an automatic restoration of rotational symmetry to some extent, as evident from the images.

\begin{figure}
    \centering
    \includegraphics[width=0.48\textwidth]{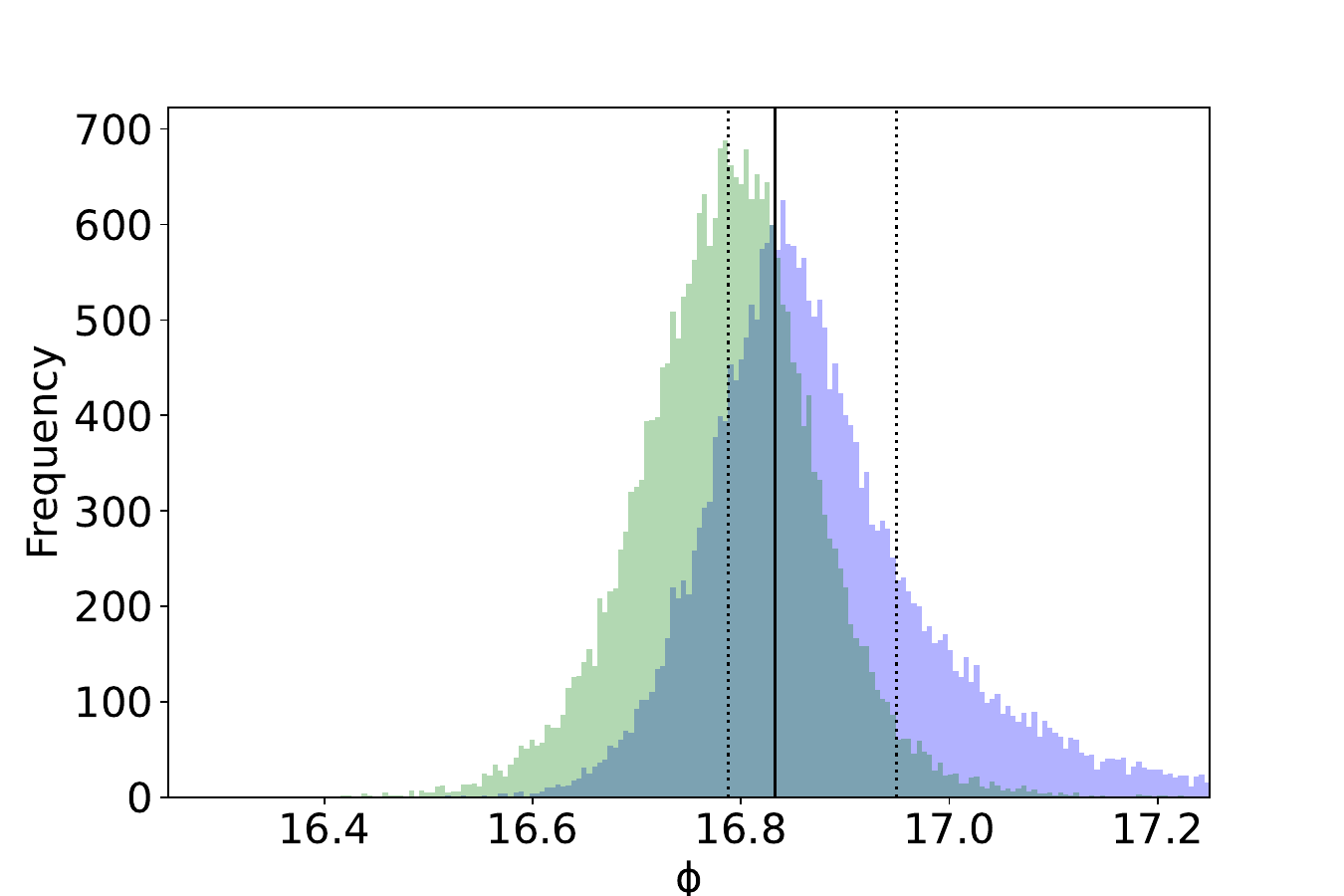}
    \caption{The frequency histograms of distributions $p_\to(\phi)$ and $p_\gets(-\phi)$, represented in blue and green respectively. The solid line in the graph represents the TFEP estimation of free energy, while the two dashed lines on its left and right sides indicate the lower and upper bound obtained from variational free energy estimations. It can be observed that the intersection of the histograms precisely aligns with the solid line.}
    \label{fig:hist}
\end{figure}

Upon a well-trained distribution $\rho_{A^\prime}(x)$, the estimations of the bounds as well as the distributions $p_\to(\phi)$ and $p_\gets(-\phi)$ are locally displayed in figure~\ref{fig:hist}. Evidently, the actual value of free energy is bounded by $\left\langle - \Phi_\gets \right\rangle _B$ and $\left\langle \Phi_\to \right\rangle _A$. Furthermore, from the distribution perspective, the two distributions $p_\to(\phi)$ and $p_\gets(-\phi)$ exhibit a certain degree of overlap, with the abscissa of the intersection point aligning with the TFEP estimation of free energy. These two observations align precisely with the theoretical framework expounded in section~\ref{sec:fe}.

Overall, with the classical Coulomb gas in a harmonic trap as an exemplar, we have demonstrated the utilization of sample data to estimate the free energy through the method of flow matching. In this approach, the estimation of free energy has attained explicit upper and lower bound, and we can also estimate the actual value of free energy from the overlap of the distribution plots. Moreover, a method for obtaining tighter bounds is demonstrated in Appendix~\ref{sec:tb}.

\section{Discussion}
\label{sec:disc}

Machine learning and physics research share some commonalities in their methods and objectives~\cite{carleo2019machine}. Both disciplines focus on the process of gathering and analyzing data to devise models capable of predicting the behavior of intricate systems. However, in practical applications, we typically utilize training and testing sets to assess the generalization performance of models. Yet, this approach merely provides an estimation of the model's performance on a given dataset, without offering an exact estimation of the error on unknown data. Similarly, when employing the variational principle to solve problems, it is also challenging to obtain an exact error estimation for the predicted results. In this context, we present a scenario where deep learning can be employed to provide upper and lower bounds for free energy.

This methodology can be applied in more practical systems in the future, such as the field of molecular science. In intricate systems, the efficacy and precision of flow matching can be further leveraged to a greater extent. Also, combined with a direct generalization on the Jarzynski equality~\cite{tang2015work}, the method may potentially have wider impacts on non-equilibrium dynamics. In the experiments conducted in this paper, we did not impose constraints on spatial symmetry. Although the experimental results indicate the restoration of spatial symmetry, in larger and more intricate systems, utilizing vector fields that consider spatial symmetry would likely have a greater impact on the training outcomes.

\section{Acknowledgments}

We thank Han Wang and Linfeng Zhang for the discussion. This project is supported by the Strategic Priority Research Program of Chinese Academy of Sciences under Grants No. XDB0500000 and No. XDB30000000, and National Natural Science Foundation of China under Grants No. 92270107, No. 12188101, No. 12122103, No. T2225018, and No. T2121001.

\bibliography{paper}

\appendix

\section{Proof of Equation~\ref{eq:vfe}}

To make the proof clearer, we will denote the samples from state $A$ as $z$ and the samples from state $A^\prime$ as $x$. 

We begin with the Kullback-Leibler divergence between the distributions of state $A^\prime$ and $B$:
\begin{equation*}
    \int^\infty_{-\infty} \rho_{A^\prime}(x)\mathrm{ln} \left( \frac{\rho_{A^\prime}(x)}{\rho_B(x)}\right) dx \geq 0.
\end{equation*}
Subsequently, by effecting a change of variables, we substitute $x$ with $z$:
\begin{equation*}
    \int^\infty_{-\infty} \rho_{A}(z) |\frac{\partial z}{ \partial x}| \left(\mathrm{ln} \left( \rho_{A}(z) |\frac{\partial z}{ \partial x}| \right) - \mathrm{ln} \left( \rho_B(x) \right) \right) |\frac{\partial x}{ \partial z}| dz \geq 0.
\end{equation*}
Simplification follows as we realize the Jacobian determinant's absolute value, $|\frac{\partial x}{ \partial z}|$, appears in both terms and consequently cancels out:
\begin{equation*}
    \int^\infty_{-\infty} \rho_{A}(z) \left(\mathrm{ln}\rho_{A}(z) - \mathrm{ln}|\frac{\partial z}{ \partial x}| - \mathrm{ln} \left( \rho_B(x) \right) \right) dz \geq 0.
\end{equation*}
By expressing the probability density using the Boltzmann distribution, we derive:
\begin{equation*}
    \int^\infty_{-\infty} \rho_{A}(z) \left( - \beta H_A(z) - \mathrm{ln}|\frac{\partial z}{ \partial x}| + \beta H_B(x) \right) dz \geq \mathrm{ln}Z_B - \mathrm{ln}Z_A,
\end{equation*}
and based on the definitions of $\Delta F$ and $\Phi_\to$, Equation~\ref{eq:vfe} can be deduced.

\begin{figure}
    \centering
    \includegraphics[width=0.48\textwidth]{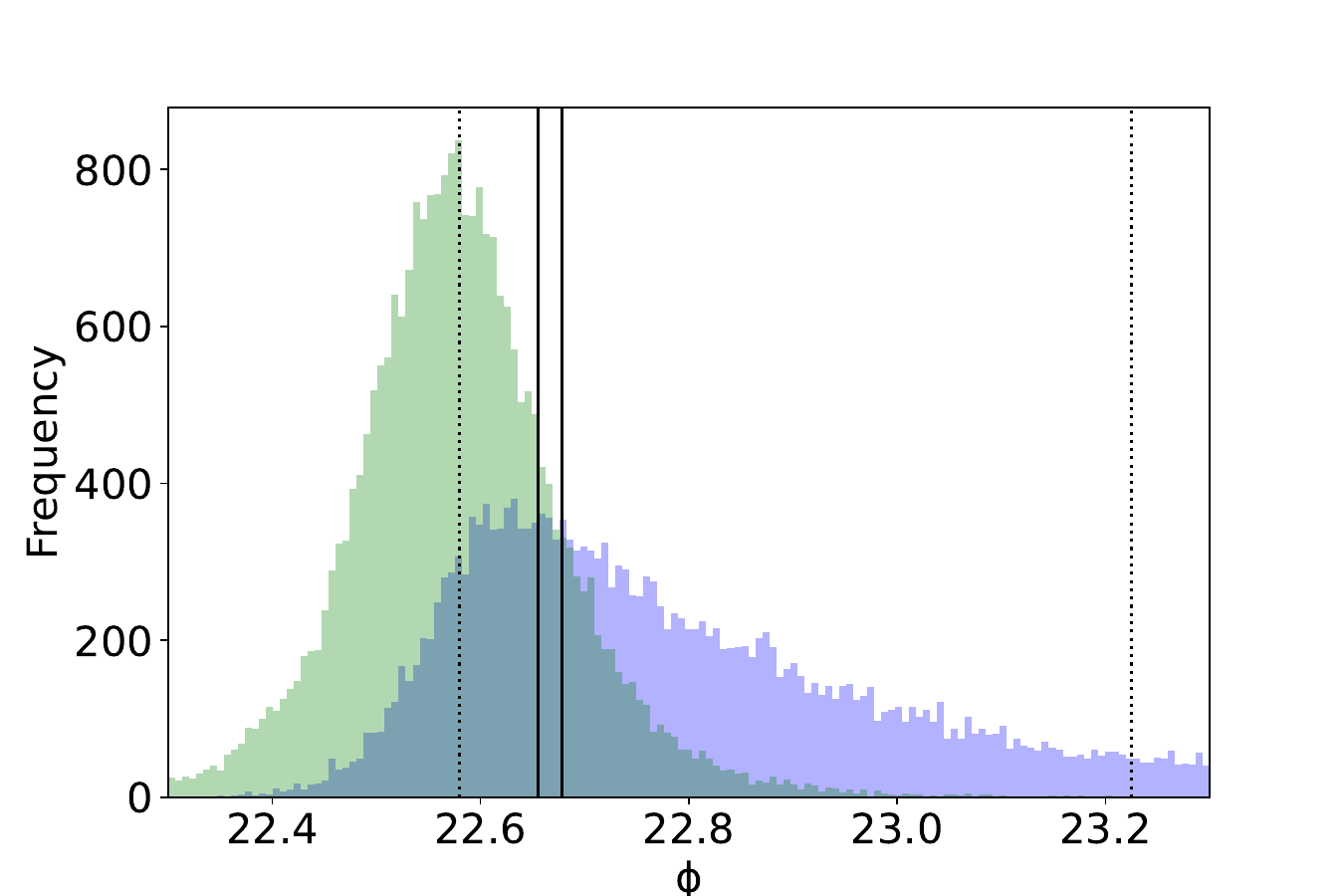}
    \caption{The frequency histograms of distributions $p_\to(\phi)$ and $p_\gets(-\phi)$, represented in blue and green respectively. The solid lines represent the lower and upper bound obtained from the ensemble averages of TFEP estimations, while the two dashed lines indicate the lower and upper bound obtained from variational free energy estimations.}
    \label{fig:tigher}
\end{figure}

\section{The Tighter Bound}
\label{sec:tb}

Besides the bound shown in Equation~\ref{eq:bound}, a tighter bound has been demonstrated~\cite{hahn2009using}
\begin{equation}
    \label{eq:tb}
    \left\langle - \Phi_\gets \right\rangle _B
    \leq \left\langle \widehat{ -\Phi_\gets} \right\rangle _B
    \leq \Delta F
    \leq \left\langle \widehat{\Phi_\to} \right\rangle _A
    \leq \left\langle \Phi_\to \right\rangle _A,
\end{equation}
where
\begin{equation}
    \label{eq:ff}
    \widehat{\Phi_\to} = -\frac{1}{\beta}\mathrm{ln}\overline{\mathrm{e}^{-\beta \Phi_\to}}
\end{equation}
and
\begin{equation}
    \label{eq:fr}
    \widehat{ - \Phi_\gets} = \frac{1}{\beta}\mathrm{ln}\overline{\mathrm{e}^{\beta \Phi_\gets}}
\end{equation}
are the TFEP estimators of the forward and reverse processes. The overline signifies taking the average of a finite set of samples $\{\Phi_i^k\}$ where $i$ denotes "forward work" $\to$ and "reverse work" $\gets$ respectively and $k$ is the number of samples. For an ensemble composed of $\{\Phi_i^k\}$, the TFEP estimator has an ensemble average
\begin{equation}
    \left\langle \widehat{\mp \Phi_i} \right\rangle _i = \mp \frac{1}{\beta} \left\langle \mathrm{ln}\overline{\mathrm{e}^{\mp \beta \Phi_i}} \right\rangle _i.
\end{equation}
Applying Jensen’s inequality to the averages of the logarithms
\begin{equation}
    \left\langle \mathrm{ln}\overline{\mathrm{e}^{\mp \beta \Phi_i}} \right\rangle _i
    \leq \mathrm{ln} \left\langle \overline{\mathrm{e}^{\mp \beta \Phi_i}} \right\rangle _i
    = \mp \beta \Delta F,
\end{equation}
we get the tighter bound shown in Equation~\ref{eq:tb}.

Equation~\ref{eq:tb} informs us that TFEP estimators are biased estimations, nevertheless, their average values can provide narrower upper and lower bounds than the variational way. When the sample size within Equation~\ref{eq:ff} and ~\ref{eq:fr} is 1, the tight bound of Equation~\ref{eq:tb} is equivalent to the bound shown in Equation~\ref{eq:bound}.

In order to make the tighter bound more pronounced on the histogram, we computed the quantities in Equation~\ref{eq:tb} on a physical system with a particle count of 7. The results are shown in Figure~\ref{fig:tigher}. In the experiment, we chose to perform the calculations on a sample set with a batch size of 25,600. It should be noted that the ensemble average of TFEP estimations $\left\langle \widehat{ -\Phi_\gets} \right\rangle _B$ and $\left\langle \widehat{\Phi_\to} \right\rangle _A$ requires averaging twice, so we divided the sample set into 100 subsets, with each subset containing 256 samples. As we can see, under the same invertible mapping $\mathcal{M}$ and the same sample set, $\left\langle \widehat{ -\Phi_\gets} \right\rangle _B$ and $\left\langle \widehat{\Phi_\to} \right\rangle _A$ bound free energy narrower significantly.

\end{document}